# Globalization Process in Emerging Capital Markets
# -- Lessons and Implications to China

**Zichong Li*** Rensselaer Polytechnic Institute

**Pengyu Huang** Renmin University of China

**Oct 1 2016**

*Undergraduate in Mathematics and Economics, Rensselaer, 110 8th Street, Troy, NY 12180, USA. Phone 1 518 368 3695, email liz19@rpi.edu.

Acknowledgement: This article is sponsored by Shanghai Stock Exchange.

# Globalization Process in Emerging Capital Markets -- Lessons and Implications to China

## Abstract

Since 2002 when China first introduced QFII (Qualified Foreign Institutional Investors) system, QFII has been developing in China for 14 years, during when RQFII, Shanghai-Hongkong Stock Connect Program, Shanghai-London Stock Connect Program furthur broadened the avenue for foreign capital to invest in Chinese Security Market. As FTA (Free Trade Area) Financial Reform Program emerged, RMB (CNY) Capital Project is likely to make the currency exchangeable. With the success in QFII, RQFII and Shanghai-Hongkong Stock Connect Program, China's long term advantage in interest rate, and the relatively low stock index value after the recent stock market crashes in mid 2015 and early 2016, foreign capitals' demand for Chinese market to loosen its restrictions continually increases. This article picks the three most representative emerging capital markets in the world, namely Taiwan, Korea and India, by comparing and analyzing their paths of globalization, attempts to shed light on China's next steps regarding globalization.

# 1. Gradual Procedure of Globalization

## 1.1 India's Steps

Although Indian capital market has a long history, before 1990s, its transaction settlement structure was primitive. Pushed by the government and regulator, National Stock Exchange of India(NSE) was founded, signaling the modernization of India's security market. By 1991 Indian market was closed with rare cross-boarder capital movements. In 1992 it opened to foreign institutional investors (FII) who had to apply through Security and Exchange Board of India (SEBI) for the qualification, at the same time foreign individual investors and general corporates who wanted to access Indian market had to apply for FII Subaccount qualification which had capital requirement of $50 million. Afterwards Indian market regulator gradually loosens these restrictions, and FII had a major impact in Indian Security Market: as of the end of 2015, FII has approximately 40% quota of the exchange-listed companies in circulation. In 2012, affected by negative factors including European Debt Crisis, dampen national growth, Rupee depreciation, India stock market didn't perform well at that time, the "barometer" index SENSEX once dropped below 16000. To attract foreign investors into the market, in 2013 India

opened its market to foreign individuals and general corporates, canceled its net asset value requirement, and greatly simpified its approval process: one only needed to go to the qualified saving institutions to open relevant accounts.

### 1.2 Korea's Steps

Korea's process of globalization is the quickest among all 3 markets. The opening of Korea Stock Exchange in 1956 signaled the formation of a modern market. Before 1980, Korean capital market was closed. Korean government issued Korea International Trust Fund and Korea Trust Fund in 1981, and three other international trust funds in Europe and Southeast Asia in 1983 where foreign investors could purchase fund shares to invest in Korean market. In 1992 Korea allowed foreign institutional investors and in succesion, individuals and general corporates to directly invest in the domestic stock market. Korea then opened its derivatives market in 1993, and bond market in 1997 to foreign investors.

### 1.3 Taiwan's Steps

Taiwan's security market was embryonic in 1950s. In 1962, the opening of Taiwan Stock Exchange signaled the formation of Taiwan security market. In 1980s Taiwan market started to globalize: in 1982 Taiwan government formulated its market plans for foreign investors, in May 1983 it allowed foreign capital to indirectly invest in Taiwan stocks, and issued overseas fund to allow foreign investors to invest in Taiwan stock market. In Dec 28, 1990, Taiwan government enacted this system allowing foreign institutional investors who satisfy certain requirements to apply to directly invest in Taiwan stock market from 1991 on. Afterwards Taiwan government continually loosened the QFII requirements and allowed more and more investment options. Finally in 1996, Taiwan government allowed general foreign individual investors to indirectly invest in the market; and in 2003, it abolished the QFII system and completely opened its security market.

## 2. Main Mechanisms of Market Globalization

### 2.1 Investor Eligibility

In early days of internationalizing, to avoid financial risk and aid market development, financial regulators usually set high entry bars to filter out good long term foreign investors, then gradually lower the entry bars to pave the way for the later complete opening. Note that Taiwan and India set high entry bars, and Taiwan and Korea's globalization process are the quickest.

Taiwan initially implemented the process through closed mutual funds, issued by domestic security companies, which allowed foreign investors to indirectly invest into Taiwan market through purchasing income certificates. Taiwan government issued QFII system on Dec 28, 1990 that allows foreign professional institutional investors to apply to directly invest in Taiwan stock market since 1991. The initial approved institutions must satisfy the following requirements: 1) Foreign-funded bank has to lie in the top 50 ranked banks in the western world with a net asset value above $300 million. 2) Foreign-funded insurance company has to be in business for more than 10 years, with a net asset value above $500 million. 3) Asset management company has to be founded for more than 5 years and currently running more than $500 million total of assets. Taiwan opened its market to foreign security companies in 1993, to mutual funds and trust funds and to individual investors in 1996, reduced the capital requirements 2 times in 1995 and 2001, and finally abolished the QFII system and completely opened the market in 2003.

Securities and Exchange Board of India (SEBI) didn't have rigid rules in net asset value, but the foreign institutional investors had to satisfy the following requirements: 1) Competency, financial stability, good reputation. 2) Regulated by its local financial regulators such that India's SEC could judge on a more subjective basis. Different from Taiwan, India allowed individuals and general corporates to invest in its markets at the very start: investors registered their subaccounts through SEBI which has a NAV (net asset value) requirement of $50 million and a recheck of eligibility every 5 years. In 2006 it allowed reinsurance company and investment consulting firms to be applicants of FII. In 2008 it furthur loosened FII application restrictions: even unregulated university funds, trust funds, charity funds and sovereign funds could also apply for FII. In 2013, India canceled the entry bars of individuals and general corporates: they just needed to open relevant accounts in India's local brokerage firms to invest.

Korean capital market had the fastest development. In early 1980s Korea allowed foreign investors to indirectly invest through international trust funds managed by Korean security companies. These funds mainly targeted foreign small and medium investors, although the amount of investment collected was limited, but these funds pave the way to furthur market opening. Korea established Korea International Trust Fund and Korea Trust Fund in 1981, and another three international trust funds in Europe and Southeast Asia in 1983. These funds invested in the Korean market with an approximate ratio: 50% in unlisted companies, 40% in listed common stocks, and 10% in other liquid securities. Korea opened its market to financial institutions, general corporates and individual investors without an entry bar since as early as 1992.

## 2.2 Investment Restrictions

Each emerging capital market has respective restrictions to foreign institutional and individual investors regarding the stock portfolio ratio and investment targets. For shareholding ratio, Taiwan and Korea initially both had strict restriction which later gradually loosened while India had strict ratio restriction all the time.

Taiwan initially stipulated that each foreign institutional investor can't own more than 5% of any specific listed company's total share; the net share of all foreign institutional investors can't exceed 10% of any specific company; each investor can't invest more than $5 to 50 million depending on specific situations while the whole market has an upper limit of $2.5 billion to all foreign institutional investors. Afterwards Taiwan altered this cap on each institution and on all foreign institutions each year, specifically in 2000 the individual upper bound was increased to a large $2 billion already. Entry bar and investment restrictions were always on the trend of gradual loosening. In the case of individual investors and general corporates, Taiwan's SEC, starting from 1996, allowed foreign individuals and general corporates to directly invest Taiwan security market. Initially the net investment by foreign investors on any specific company can't exceed 20% of its total circulated shares;

foreign investors living in Taiwan had no restrictions while other foreign natural persons had investment upper limit of $5 million and legal persons had upper limit of $20 million. In 1996 Taiwan raised the cap of all foreign natural persons regarding any specific company to 30% of its total shares, and 15% for each individual. Later these two ratios both increased to 50%.

Starting from 1992, Korea opened its security market directly to foreign institutions, individuals, and general corporates, ruling that each separate investor and all foreign investors can't own respectively more than 3% and 10% (8% if public company) of any company's total shares. Financial Crisis in 1997 made Korean government realize the importance of attracting foreign capital. In December 1997, foreign institutions and individuals' ratio caps of investing Korean stock increased respectively to 50% and 55%. In May 1998, all restrictions regarding investment limits, with the exception on a few public legal persons, were completely abolished. At the end of 2000, except special restrictions on fields including telecommunication, broadcasting and aviation, Korean lifted almost all restrictions regarding foreign capital involvement. Due to this reason foreign investments in Korea grew rapidly.

Concurrently, India still had severe restrictions on FII investments regarding shareholding ratio: any separate FII account can't own more than 10% of any company; any separate FII subaccount can't own more than 5% of any company; all foreign investors combined can't own more than 49% (24% in certain fields) of any company. In 1996 the FII subaccount upper limit was lifted from 5% to 10%.

## 2.3 Capital Control Measures

The purpose of capital control measures is to avoid foreign high-frequency purchase and sale of stocks and frequent capital inflows and outflows that could cause fluctuation in security market and foreign exchange market, with an intent to aid foreign capitals towards long term investments and to stabilize its security and foreign exchange market. There were mainly two modes of regulation: rules regarding capital inflow and outflow, and tax regulation.

Taiwan mainly adopted restrictions of capital inflow and outflow to regulate its market. It's worth noting that Taiwan takes into account the international variation of capital movements and adjust these restrictions counter-cyclically. Initially in 1991, Taiwan's foreign exchange admission stipulated that foreign investors, after approval, had to remit in the principal in 3 months, and can only be remitted out after another 3 months of investment, at most once per year. Taiwan changed the remit-in time cap to 6, 3, 4 months respectively in 1991, 1993 and later 1993, and canceled this restriction in January 1996. In 1997 Southeast Asia

Financial Crisis emerged, to encourage QFII capital inflow, Taiwan allowed the remit-in funds below $50 million to be free of foreign exchange administrative department‘s approval, and extended the remit-in deadline to one year, which increased to two years in 2001.

Korea and India mainly adapted tax measures to regulate foreign capital movement. Comparatively Korea had an easier system which was mainly based on tax, with the standard amount of minimum of 27% of capital gains or 11% of total securities; dividend income was taxed 25%. On the other hand, India levied different tax ratios based on the fund's duration in India: if the fund leaves in less than 1 year, then it's taxed 10% on capital gains and 20% on dividend and interests; otherwise only 10% tax on capital gains would be applied.

## 2.4 Information Disclosure Requirements

Among the three emerging markets, Taiwan and India had stricter information disclosure requirements. For India, when one applies for FII account or subaccount, he needs to fill Form-1 with lots of details which would be made public, both for individuals and institutions. SEBI is responsible for the examination process while the application information also has to be sent to the Reserve Bank of India (RBI). Every day, India would disclose its daily FII net purchase and sale amounts and add it to the yearly data. For shareholding ratio, when FII and their subaccounts own a net of more than 24% or 49$ of any company's total shares, it needs to be approved by RBI. At the same time, Taiwan also requires QFII's custodian banks to report daily inflow and outflow amounts and stock portfolios to Taiwan's SEC and central bank; specifically if any foreign investor owns more than 10% of any company's total shares, it needs to be reported to SEC immediately and disclosed to the public. SEC would also disclose QFII's purchase and sale of each stock and its market shares to the public.

# 3. Timing of Market Globalization

## 3.1 Macroeconomics

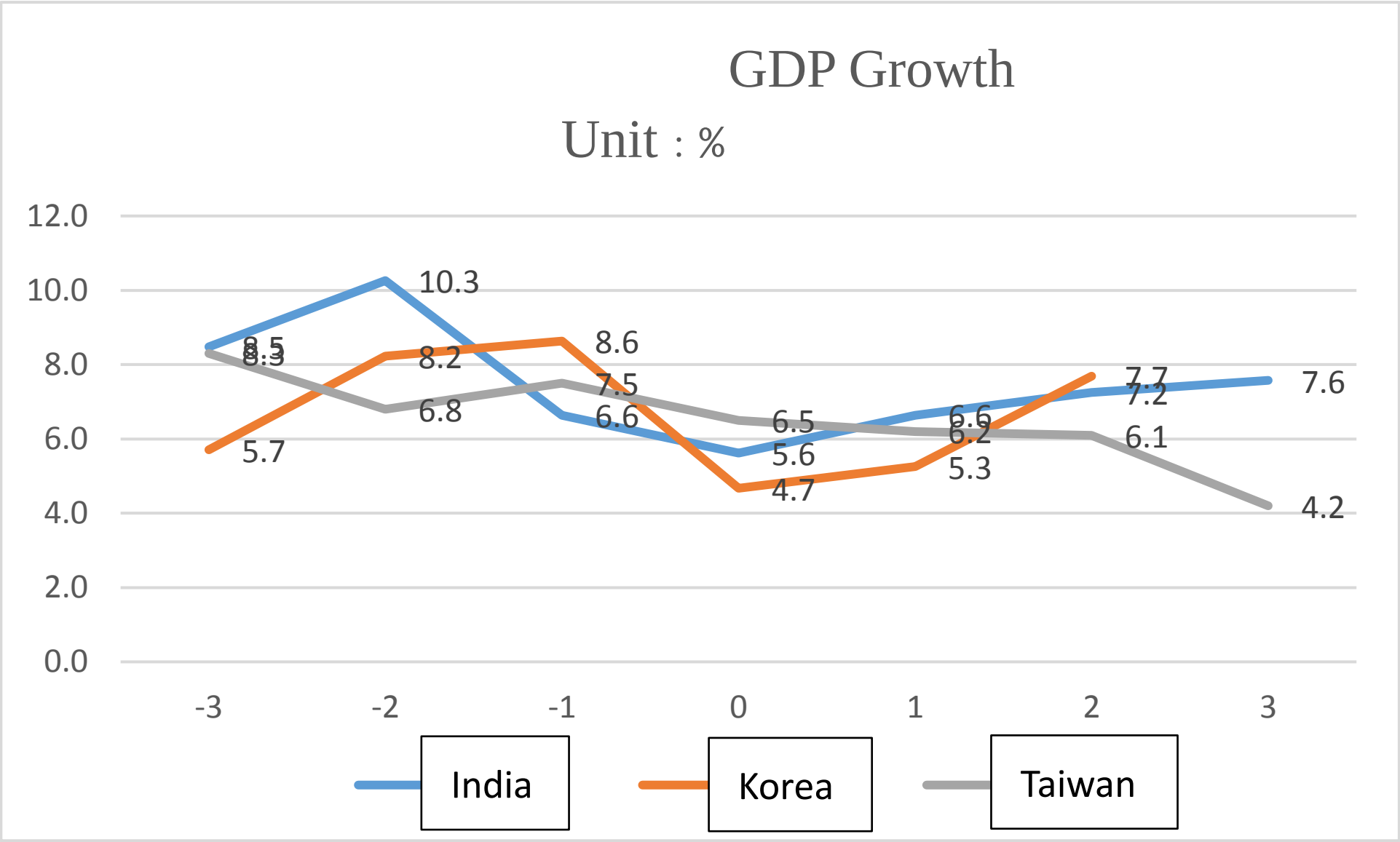

Note: Horizontal axis is the relative year, with 0, the basis year, as the year when direct investment

is first allowed, other years refer to this basis year.

According to macroeconomic data from each region, pick GDP Growth Rate as reference, each region's decision to open up its market to foreign investors collide with its economic downturn. India, Korea and Taiwan had GDP Growth Rate of 5.6%, 4.7% and 5.6% respectively, which was on the lower side of the historical values. Choices of completely opening up the security market are usually motivated by pressure of economic downturn: lack of inner growth forces the government to consider importing more foreign money into the security market to facilitate quick recovery of the country's real economy. In fact, each region had substantial economic recovery after completely opening up its market.

### 3.2 Stock Market Structure

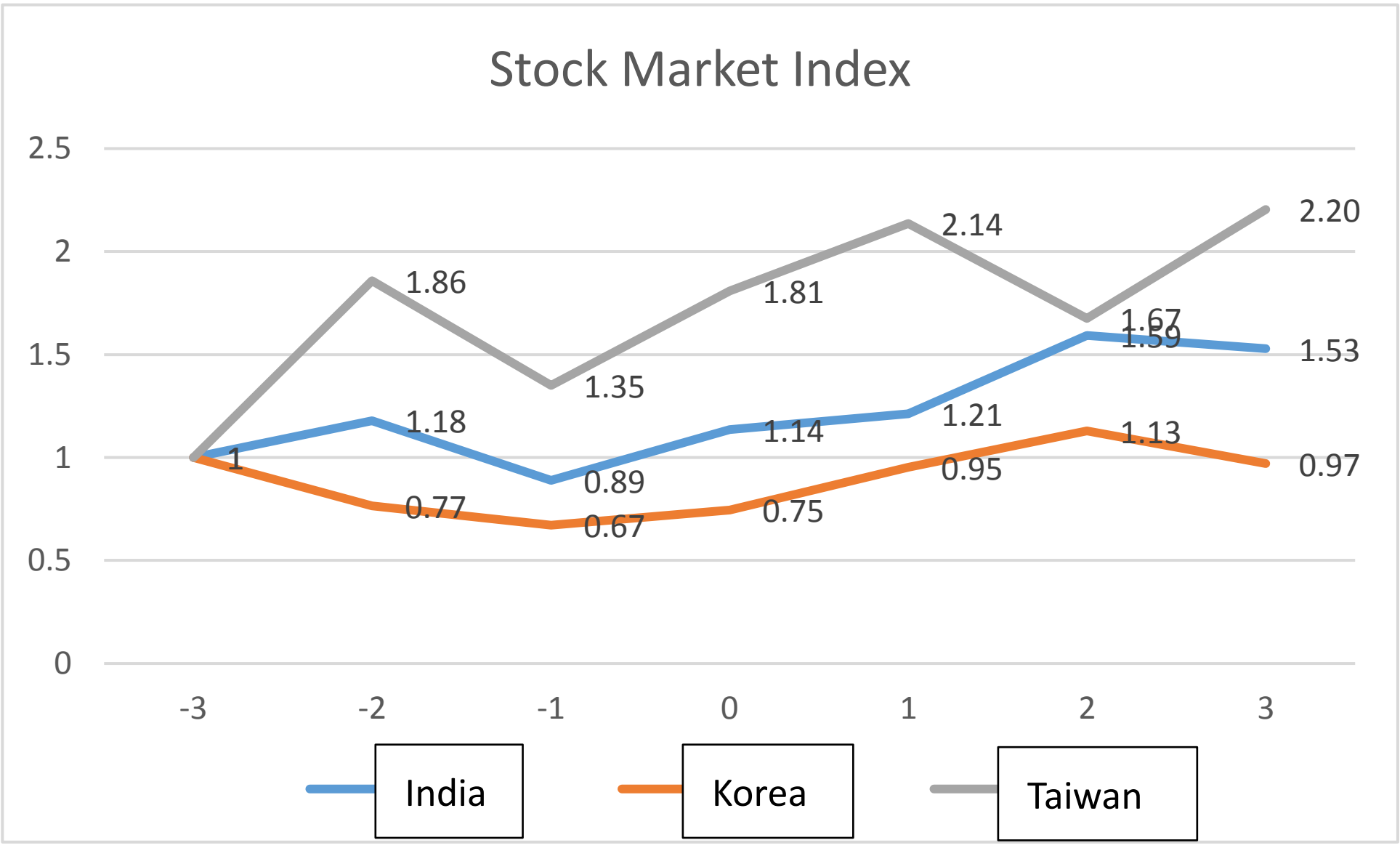

Note: Horizontal axis is the relative year, with 0, the basis year, as the year when direct investment is first allowed, other years refer to this basis year. Index is converted to 1 at year -3.

According to the stock market indexes, before each region decided to open up its market, its stock index usually had a low value. This fact can be attributed to government's wish to import foreign capital to recover the stock market, or on the other side, the fact that the relatively lower valuation made foreign capitals more willing to add respective domestic stocks to their portfolios.

## 3.3 Foreign Exchange and Interest Rate System Development

According to the other emerging capital markets, loosening of account opening restrictions is usually synchronized with the foreign exchange and interest rate system to mitigate the shock of cross-boarder capital movement to the domestic economy and financial stability.

In China's perspective, Korea and China has many similarities regarding mode of economic development. From 1980 to 1999 Korea's exchange rate were regulated; after the Southeast Asia Financial Crisis in 1997, Korea had to switch to free floating exchange rate. Thus in 1992 when Korea loosens restriction on account opening, were able to realize capital convertibility 1 year afterwards, given the fact that South Korea began to liberalize most of the short-term interest rates of banks and non-bank financial institutions in November 1991, and gradually began to promote market-oriented interest rate reform, and in 1997, interest rate marketization was fully realized. LULUL In 1989, Taiwan concurrently used market-oriented exchange rate and interest rate system, thus when opened QFII in 1992, had a high degree of market-orientation. Thus only four years later in 1996, Taiwan changed to a direct account system. India achieved capital account convertibility in 2000, followed by the implementation of interest rate marketization in 2011, setting a solid foundation for its full opening of unrestricted individual direct accounts in 2013. In 2013, India canceled the $50 million requirement of the individual investors and general corporate, when opening up direct accounts had its exchange rate system fixed to the target exchange rate of dollar while its interest rate achieved marketization at the same time.

## 4. Emerging Capital Markets' Globalization Processes -- Lessons and Implications to China

The financial reform of the Free Trade Area (FTA) and the construction of Shanghai's International Financial Center (IFC) created an opportunity for our country to transition from indirect foreign investments to direct foreign investments. By summarizing and analyzing the opening process of emerging international capital markets, this paper concludes with the following lessons and implications to China, regarding the transition from indirect account opening to direct account opening.

## 4.1 Overall Planning, Step By Step Opening Procedure

China introduced the QFII system in 2003, and has now gone through 13 years, regulators and the market have both accumulated some experience regarding foreign investment and regulation. From the international scope of the globalization process, QFII system is usually followed by the allowance of direct account opening. At the same time, having experienced last year's stock market crash, the market valuation is currently at an acceptable level, and QFII quota specifically has been continuously in a state of short supply, indicating the high demand of foreign portfolios to include Chinese securities. In addition, the financial reform of FTA, the convertibility of capital account under FT account, the further marketization of exchange rate and interest rate, and the construction of international financial assets trading platform of FTA indicate that the current transition from indirect account opening to direct account opening is under a favorable time window. But before the opening the regulator should still pay attention to the orderly openings of the financial markets and the entry bars.

Compared with other emerging markets which, before implementing a direct account system, all realized capital convertibility and exchange rate and interest rate liberalization. However, China's current capital account is still not fully convertible, while exchange rate and interest rate marketization has not yet been fully realized. Therefore, in the capital market globalization process, we must give full credit to the advantages of FTA's financial policies, and use the FTA trading platform as the channel to open up exchanges and overseas funds, and take the lead in opening accounts directly.

## 4.2 Coordination with Foreign Exchange and Interest Rate

In order to make long-term funds willing to invest long in the domestic securities market, the SEC not only has to supply strict regulatory measures, but also should pay more attention to improve its financial market, i.e. exchange rate marketization and interest rate marketization. Through the improvement of financial policy, the regulator should aid to create a highly efficient and transparent financial market, so that foreign funds can contribute to the real economy and truly help the development of real enterprises.

## 4.3 Centralized Information Regulatory System

During the stock market crash last year there were indeed some foreign funds which used the imperfections of the stock market to conduct malicious shorting, resulting in a domestic panic towards foreign funds . Drawing lessons from the practices of other emerging markets to China, for personal and general corporate direct account opening, the information disclosure system should still be improved, foreign capital flows should be monitored on a real-time basis, and long-term value investment should also be encouraged to the foreign funds.